\documentclass[fleqn,10pt]{wlscirep}
\usepackage[utf8]{inputenc}
\usepackage[T1]{fontenc}

\usepackage{lipsum}
\usepackage{color}
\usepackage{algorithm}
\usepackage{algorithmic}
\usepackage{amsmath,amsfonts,amssymb}
\usepackage{subfigure}
\usepackage{lineno}
\usepackage{graphicx}
\usepackage{bm}
\usepackage{comment}
\usepackage{xurl}

\title{Global field reconstruction from sparse sensors with Voronoi tessellation-assisted deep learning}

\author[1,2,*]{Kai Fukami}
\author[3]{Romit Maulik}
\author[4]{Nesar Ramachandra}
\author[2]{\\Koji Fukagata}
\author[1]{Kunihiko Taira}
\affil[1]{Department of Mechanical and Aerospace Engineering, University of California, Los Angeles}
\affil[2]{Department of Mechanical Engineering, Keio University}
\affil[3]{Argonne Leadership Computing Facility, Argonne National Laboratory}
\affil[4]{High Energy Physics Division, Argonne National Laboratory}

\affil[*]{kfukami1@g.ucla.edu}

\begin{abstract}
Achieving accurate and robust global situational awareness of a complex time-evolving field from a limited number of sensors has been a longstanding challenge.  
This reconstruction problem is especially difficult when sensors are sparsely positioned in a seemingly random or unorganized manner, which is often encountered in a range of scientific and engineering problems.  
Moreover, these sensors can be in motion and can become online or offline over time.  
The key leverage in addressing this scientific issue is the wealth of data accumulated from the sensors.
As a solution to this problem, we propose a data-driven spatial field recovery technique founded on a structured grid-based deep-learning approach for arbitrary positioned sensors of any numbers.
It should be noted that the na\"ive use of machine learning becomes prohibitively expensive for global field reconstruction and is furthermore not adaptable to an arbitrary number of sensors.  
In the present work, we consider the use of Voronoi tessellation to obtain a structured-grid representation from sensor locations enabling the computationally tractable use of convolutional neural networks.
One of the central features of the present method is its compatibility with deep-learning based super-resolution reconstruction techniques for structured sensor data that are established for image processing.  
The proposed reconstruction technique is demonstrated for unsteady wake flow, geophysical data, and three-dimensional turbulence.  
The current framework is able to handle an arbitrary number of moving sensors, and thereby overcomes a major limitation with existing reconstruction methods.
The presented technique opens a new pathway towards the practical use of neural networks for real-time global field estimation.
\end{abstract}
\begin{document}

\flushbottom
\maketitle
\thispagestyle{empty}

\section{Introduction}

Spatial field reconstruction from limited local sensor information is a major challenge in the analysis, estimation, control, and design of high-dimensional complex physical systems.
For complex physics including geophysics~\cite{manohar2018data}, astrophysics~\cite{akiyama2019first}, atmospheric science~\cite{alonso2010novel,mishra2014compressed}, and fluid dynamics~\cite{FFT2020}, traditional linear theory-based tools, {including Galerkin transforms~\cite{boisson2011three,noack1994global}, linear stochastic estimation~\cite{adrian1988stochastic,SH2017} and Gappy proper orthogonal decomposition~\cite{BDW2004},}
have faced challenges in reconstructing global fields from a limited number of sensors.
Neural networks have emerged as hopeful nonlinear alternatives to reconstruct chaotic data from sparse measurements in an efficient manner~\cite{lecun2015deep,BNK2020}.
However, there are key limitations associated with neural networks for field reconstruction. 
One of the biggest difficulties is the applicability of neural network-based methods to unstructured grid data.
Almost all practical experimental measurements or numerical simulations rely on unstructured grids or non-uniform/random sensor placements.
These grids are not compatible with convolutional neural network (CNN)-based methods which are founded on training data being \emph{structured} and \emph{uniformly} arranged~\cite{LBBH1998,FFT2019a}.
While a multi-layer perceptron (MLP)~\cite{RHW1986} can handle unstructured data, its use is sometimes impractical due to their globally connected network structure.
Moreover, MLPs cannot handle sensors that may go offline or move in space.
On the other hand, graph convolutional networks (GCNs) have been utilized to perform convolutions on unstructured data~\cite{wu2020comprehensive}. 
However, GCNs are also known to scale poorly and their state-of-the-art applications have been limited to the order of $10^5$ degrees of freedom~\cite{mallick2020transfer}.
Even such applications of GCNs have required distributed learning on several hardware accelerators. 
This restricts their utilities for practical field reconstructions. 
A greater limitation stems from the fact that all methods fail to handle spatially moving sensors. 
This implies that applications of these conventional tools are limited to a fixed sensor arrangement as that used in a training process. 
This limitation is a major hindrance to practical use of these reconstruction techniques, since experimental sensor locations commonly evolve over time. 
The framework that integrates convolutional architectures with time-varying unstructured data is crucial for bridging the gap between structured field reconstructions and practical problems~\cite{chai2020deep}.

In response to the aforementioned challenges, we propose a method that incorporates sparse sensor data into a CNN by approximating the local information onto a structured representation, while retaining the information of spatial sensor locations. 
This is achieved by constructing a Voronoi tessellation of the unstructured data set and adding the input data field corresponding to spatial sensor locations through a mask. 
Voronoi tessellation projects local sensor information onto the structured-grid field based on the Euclidean distance.
The currently technique achieves accurate field reconstructions from arbitrary sensor locations and varying numbers of sensors with existing CNN architectures.
The present formulation will impact a wide range of research fields that rely on fusing information from discrete sensors{, e.g, buoy-based sensors and tracers in particle tracking velocimetry~\cite{machicoane2019recent}.}

\section{Problem setup and approach}
\label{sec:method}

\begin{figure*}
    \centering
    \includegraphics[width=1\textwidth]{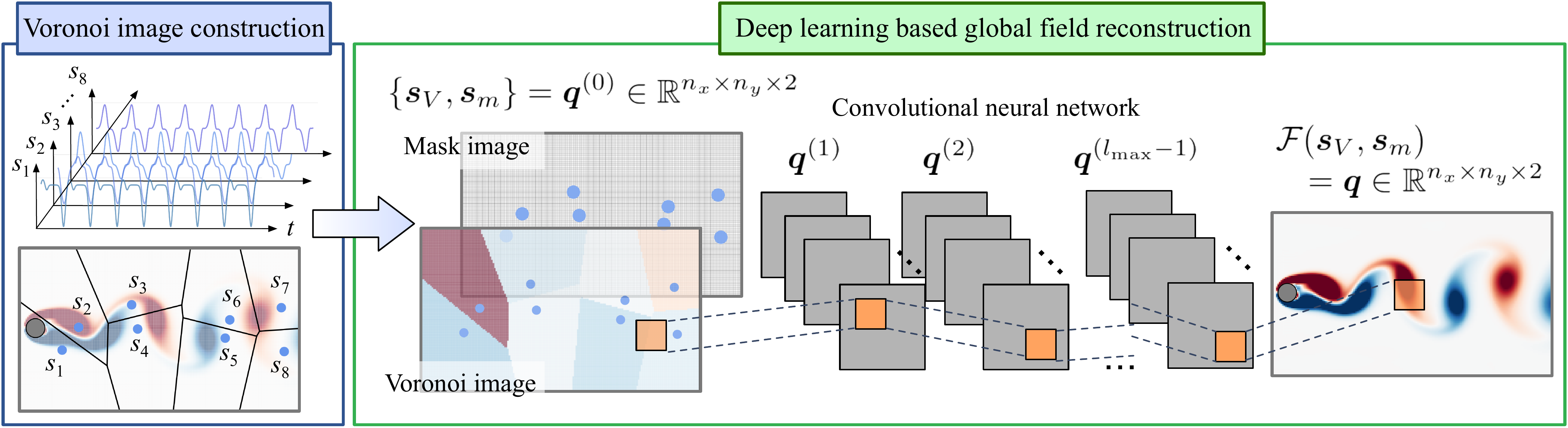}
    \caption{
    Voronoi tessellation aided global data recovery from discrete sensor locations for a two-dimensional cylinder wake. Input Voronoi image is constructed from 8 sensors. The Voronoi image is then fed into a convolutional neural network with the mask image. In the mask image, a grid having a sensor (blue circle) has 1 while otherwise 0.}
    \label{fig1}
\end{figure*}

Our objective is to reconstruct a two-dimensional global field variable ${\bm q} \in \mathbb{R}^{n_x \times n_y}$ from local sensor measurements ${\bm s} \in \mathbb{R}^n$ at locations ${{\bm x}_{\bm s}}_i \in \mathbb{R}^2, i=1,\ldots,n$.
Here, $n_x$ and $n_y$ respectively denote the number of grid points in the horizontal and vertical directions on a high-resolution field, and $n$ indicates the number of local sensor measurements.
The challenge here is to handle arbitrary numbers of sensors at any locations over the field.
The number of sensors can be changed in time and the sensors can be moving.
The reconstruction process should be performed with only a single machine learning model to avoid retraining when sensors move or change their numbers. 

{To achieve the goal of the present study, we utilize two input data files (images) for the CNN:
\begin{enumerate}
    \item Local sensor measurements projected on Voronoi tessellation ${\bm s}_V={\bm s}_V({\bm s})\in \mathbb{R}^{n_x \times n_y}$.
    \item Mask image ${\bm s}_m={\bm s}_m(\{{{\bm x}_{\bm s}}_i\}_{i=1}^{n}) \in \mathbb{R}^{n_x \times n_y}$ which contains the local sensors positions, defined as
    \[
  {\bm s}_m(\{{{\bm x}_{\bm s}}_i\}_{i=1}^{n}) = \begin{cases}
    1 & \text{if ${\bm x}={{\bm x}_{\bm s}}_i$ for any $i$,} \\
    0 & {\text{otherwise.}}
  \end{cases}
\]
\end{enumerate}
The two input images above are provided to a machine learning model ${\cal F}$ such that ${\bm q}={\cal F}({\bm s}_V,{\bm s}_m)\in \mathbb{R}^{n_x \times n_y}$, where ${\bm q}$ is the desired high-resolution field.
With these two input vectors holding magnitude and position information of the sensors, the present idea can deal with arbitrary sensor locations and arbitrary numbers of sensors.
It should be noted that reconstruction cannot be achieved with conventional methods, including MLPs and CNNs due to their structural constraints.} 
In what follows, we introduce Voronoi tessellation and machine learning framework, which are the two key components in the present approach.\footnote{Sample codes are available on \url{https://github.com/kfukami/Voronoi-CNN}.}

\subsection{Voronoi tessellation}
\label{sec:VD}

To use a machine learning framework, the sensor data needs to be projected into an image file in an appropriate manner.
Voronoi tessellation~\cite{Voronoi1908} is a simple and spatially optimal projection of local sensor measurements onto the spatial domain. 
This tessellation approach optimally partitions a given space $E$ into $n$ regions $G=\{g_1,g_2,...,g_n\}$ using boundaries determined by distances $d$ among $n$ number of sensors $s$~\cite{aurenhammer1991voronoi}.
Using a distance function $d$, Voronoi tessellation can be expressed as
\begin{eqnarray}
g_i=\{s_i\in E |d(s_i,g_i)<d(s_i,g_j),~j\neq i\}.
\end{eqnarray}
Hence, for a Euclidean domain, the Voronoi boundaries between sensors are their bisectors.

Voronoi tessellation has two important characteristics which provide the foundation for the present approximation constructed with local sensor measurements~\cite{aurenhammer1991voronoi}.
One is that each area in a Voronoi tessellation is convex.
{This property enables us to establish a Voronoi tessellation using bisections in a simple manner.}
The other is that a Voronoi tessellation does not include other sensors inside it when a circle centered at the vertex of a Voronoi region $g$ passes neighboring sensors (Empty-circle property).
{This implies that each Voronoi region $g$ is optimal for each sensor $s$ in a Voronoi tessellation.} 
{Note that the Voronoi discretization is influenced by the computational domain size.
Experimental setups are also influenced in analogous manner with their finite fields of view.}
Additional details on the mathematical theory of Voronoi tessellation can be seen in the study of Aurenhammer~\cite{aurenhammer1991voronoi}.

The spatial domains to be discretized by Voronoi tessellation and the high-resolution data are taken to be the same size.  
All grid points in each portion of the Voronoi image have its representative sensor value.
Since Voronoi tessellation provides a structured-grid representation of measurements from arbitrary placed sensors, the present approach enables us to use existing CNNs devised for structured grid data.
Note that a Voronoi tessellation needs to be performed solely once if sensors are stationary.
If the number of sensors change over time, only local regions in direct vicinity of added or removed sensors need to undergo tessellation in an adaptive manner.

\subsection{Convolutional neural network}
\label{sec:CNN}

The aforementioned Voronoi tessellation enables the use of deep learning through a structured-grid CNN~\cite{LBBH1998,FFT2019a,FFT2020b}.
Our CNN design is composed of convolutional layers as shown in figure~\ref{fig1}.
The convolutional layer extracts key features of input data through filtering operations,
\begin{equation}
    q_{ijm}^{(l)}=\psi\left(\sum^{K-1}_{k=0}\sum^{H-1}_{p=0}\sum^{H-1}_{c=0}q_{i+p-C,j+c-C,k}^{(l-1)}h_{pckm}+b_{m}\right),
\end{equation}
where {$C={\rm floor}(H/2)$,} $q_{ijm}^{(l-1)}$ and $q_{ijm}^{(l)}$ are the input and output data at layer $l$, respectively; $h_{pckm}$ represents a filter of size of $\left(H\times H\times K\right)$ and $b_m$ is the bias.
In this study, the number of channels $K$ is set to 1.
The number of layers $l_{\rm max}$, the size of filter $H$, and the number of filter $m$ are set to 9, 7, and 48, respectively, for the present study. 
The output of each filter operation is passed through an activation function $\psi$ which is chosen to be the rectified linear unit (ReLU) $\psi(z)={\rm max}(0, z)$~\cite{NH2010}.
Moreover, the ADAM optimizer~\cite{Kingma2014} is utilized with an early stopping criterion~\cite{prechelt1998} for the training process, which undergoes a three-fold cross validation~\cite{Bruntonkutz2019}.
Error are assessed using the $L_2$ norm and ensemble evaluations.
{The present idea can also be extended to three-dimensional problems without difficulty.}

{For the input to ${\cal F}$, we provide the Voronoi tessellation ${\bm s}_V \in \mathbb{R}^{n_x \times n_y}$ for sensor readings, with a mask image for input sensor placements ${\bm s}_m \in \mathbb{R}^{n_x \times n_y}$, i.e., ${\bm q}^{(0)}=\{{\bm s}_v,{\bm s}_m\}\in \mathbb{R}^{n_x \times n_y \times 2}$.
The output from the CNN is a high-resolution flow field ${\bm q}\in \mathbb{R}^{n_x \times n_y \times 1}$ that corresponds to the global field. 
In summary, the learning process of the present CNN can mathematically be formulated as 
\begin{eqnarray}
{\bm w}={\rm argmin}_{\bm w}||{\bm q}-{\cal F}(\{{\bm s}_V, {\bm s}_{m}\};{\bm w})||_2,
\label{eq:argmin}
\end{eqnarray}
where $\bm w$ denotes weights (filters) of CNN. 
We are now able to take measurements and locations by projecting them onto Voronoi tessellation ${\bm s}_V$ and the mask image ${\bm s}_m$, and reconstruct the global field variable ${\bm q}$ with deep learning ${\cal F}$.
}

\section{Applications}
\label{sec:result}

{We demonstrate the use of the present Voronoi-based CNN for global fluid flow reconstruction.
Our examples include laminar cylinder wake, geophysical data, and three-dimensional wall-bounded turbulence, which contain strong nonlinear dynamics over a wide range of spatio-temporal scales.}

\subsection{Example 1: two-dimensional cylinder wake}
\label{sec:2Dcy}

\begin{figure*}
    \centering
    \includegraphics[width=1\textwidth]{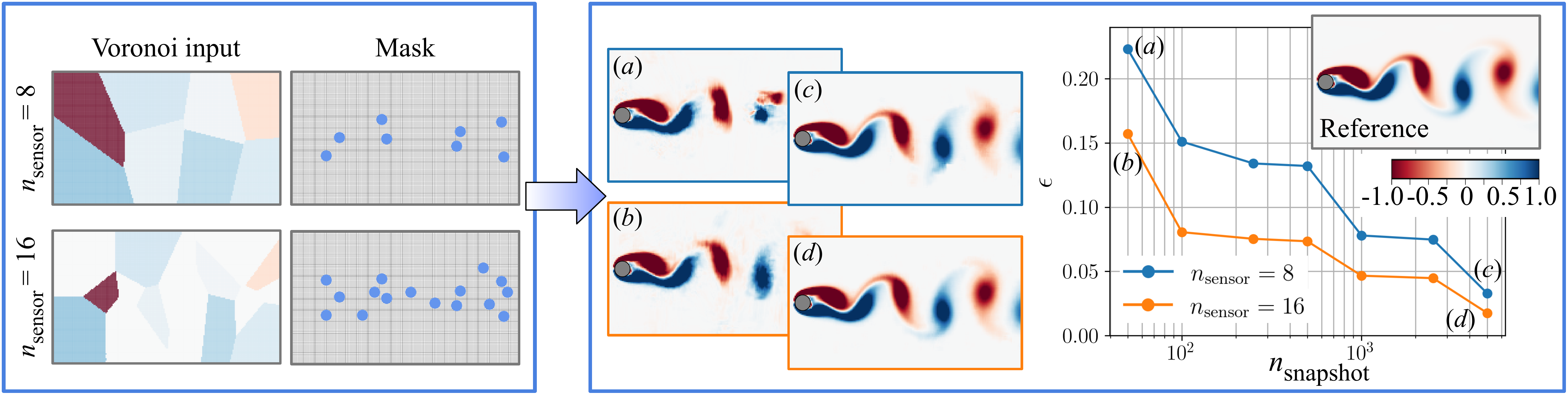}
    \caption{Voronoi-tessellation aided spatial data recovery of a two-dimensional cylinder wake with $n_{\rm sensor}=8$ and $16$. The input Voronoi image, input mask image holding sensor locations, and the reconstructed vorticity field are shown. 
    Dependence of the reconstruction ability on the number of training snapshots is also shown. {$(a)$~$\{n_{\rm sensor},n_{\rm snapshot}\}=\{8,50\}$, $(b)$~$\{n_{\rm sensor},n_{\rm snapshot}\}=\{16,50\}$, $(c)$ $\{n_{\rm sensor},n_{\rm snapshot}\}=\{8,5000\}$, and $(d)$ $\{n_{\rm sensor},n_{\rm snapshot}\}=\{16,5000\}$.} 
    }
    \label{fig4}
\end{figure*}

We first consider the Voronoi-based fluid flow reconstruction for a two-dimensional unsteady laminar cylinder wake at a diameter-based Reynolds number $Re_D=100$.
The training data set is prepared with a direct numerical simulation (DNS)~\cite{TC2007,CT2008} which numerically solves the incompressible Navier--Stokes equations.
In this study, we consider the flow field data around a cylinder body for the training and demonstration, i.e., ($(x/D)^*$, $(y/D)^*$) = $[-0.7, 15] \times [-5, 5]$ and ($N_x$, $N_y$) = (192, 112).
The vorticity field is used for both input and output attributes to the CNN model in this case. 
The training data spans approximately 4 vortex shedding periods.
We examine the dependence of the reconstruction on the amount of training data.
The number of sensors $n_{\rm sensor}$ is set to 8 and 16 with fixed input sensor locations for both training and testing.

The reconstructed fields are shown in figure~\ref{fig4}.
The reconstructed vorticity fields are in excellent agreement comparing to the reference vorticity field ${\omega}$ and in terms of the $L_2$ error norm $\epsilon=||{\omega}_{\rm ref}-{\omega}_{\rm ML}||_2/||{\omega}_{\rm ref}||_2$
{, where $\omega_{\rm ref}$ and $\omega_{\rm ML}$ denote the reference and the reconstructed vorticity fields, respectively.}
It can be seen from the reconstructed vorticity field that the vortices and shear layers in the near and far wakes are provided by the present deep learning technique with great accuracy and detail. 
The vorticity field for $n_{\rm sensor}=8$ shows some low-level reconstruction error due to the low number of sensors. 
When $n_{\rm sensor}$ is doubled to 16, we observe the error in the reconstruction is reduced by half with accurate recovery of the global flow field.
Furthermore, reasonable data recovery can be achieved with $n_{\rm sensor}=16$ using as few as 50 training snapshots.

\subsection{Example 2: NOAA sea surface temperature}
\label{sec:noaa}

Next, let us consider the NOAA sea surface temperature data collected from satellite and ship-based observations (\url{http://www.esrl.noaa.gov/psd/}).
The data is comprised of weekly observations of the sea surface temperature with a spatial resolution of $360\times180$.
We use 1040 snapshots spanning from 1981 to 2001, while the test snapshots are taken from 2001 to 2018. 
For this example, we take the sensors to be placed randomly over the water. 
The number of sensors for training is set to $n_{\rm sensor, train}=\{10, 20, 30, 50, 100\}$ with 5 different arrangements of sensor locations, amounting to 25 cases.
{The sensor locations are randomly provided for each snapshot.}
For the test data, we also consider unseen cases with 70 and 200 sensors such that $n_{\rm sensor, test}=\{10, 20, 30, 50, 70, 100, 200\}$. 
These numbers of sensors for the test data correspond to \{0.0154\%, 0.0309\%, 0.0463\%, 0.0772\%, 0.108\%, 0.154\%, 0.309\%\} against the number of grid points over the field.
{Both seen and unseen sensor locations for all numbers of sensors are considered for the present test demonstrations.}
We emphasize that only a \emph{single} machine learning model is trained and used for all combinations of sensor numbers and sensor placements.

\begin{figure*}[b]
    \centering
    \includegraphics[width=1\textwidth]{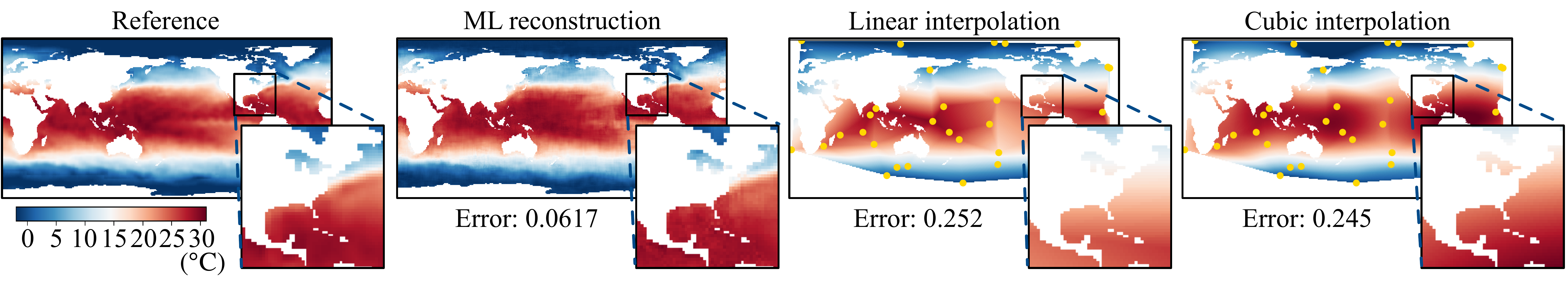}
    \caption{Comparison of spatial data recovery for NOAA sea surface temperature with $n_{\rm sensor}=30$.
    }
    \label{fig6_0}
\end{figure*}

Let us demonstrate the global sea surface temperature reconstruction in figure~\ref{fig6_0}. 
As a test case, we use a low $n_{\rm sensor}=30$. 
The reconstructed global temperature field by the present model shows great agreement with the reference data.
This figure also reports the $L_2$ error norm $\epsilon=||T_{\rm ref}-T_{\rm ML}||_2/||T_{\rm ref}||_2$, where $T_{\rm ref}$ and $T_{\rm ML}$ are respectively the reference and reconstructed temperature fields. 
We here also compare our results to standard linear and cubic interpolation.
Since those are simple methods, fine structures cannot be recovered and the $L_2$ errors are larger than that of the present method. 
These trends are noticeable from comparing the zoomed-in temperature contours.
The interpolation schemes are unable to reconstruct the fine-grained features of the temperature fields accurately.
However, the proposed technique performs very well.
In addition to enhanced reconstruction, the present method is able to recover the whole field, while classical interpolation methods cannot extrapolate beyond the convex hull covered by the sensors, as evident from figure~\ref{fig6_0}. 
This observation also speaks to the significant advantage of the present model.

\begin{figure*}[t]
    \centering
    \includegraphics[width=1\textwidth]{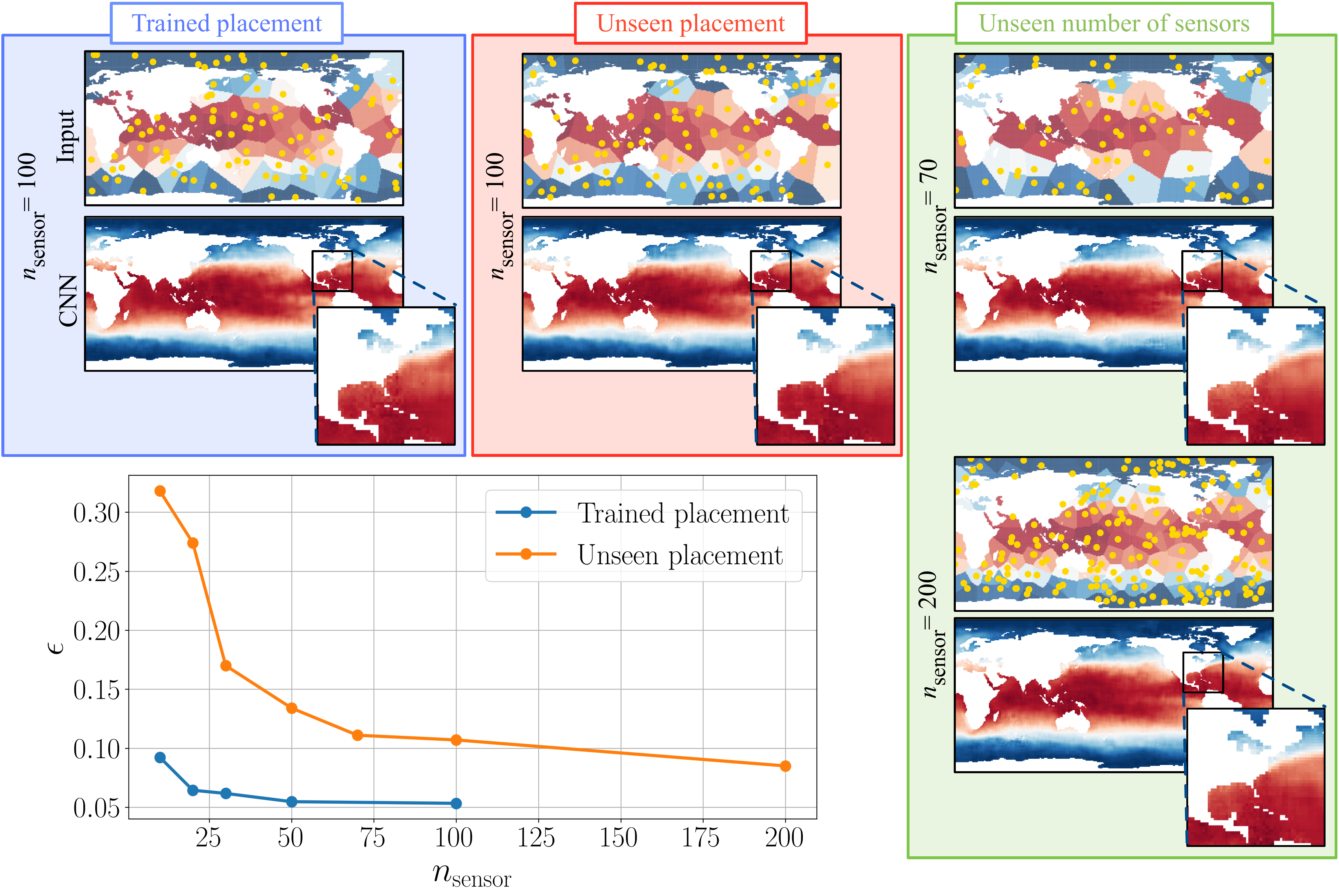}
    \caption{
    Voronoi based spatial data recovery of NOAA sea surface temperature. We show the representative reconstructed fields with $n_{\rm sensor}=100$ which corresponds to the number of sensors contained in the training data and $n_{\rm sensor}=\{70,200\}$ which correspond to cases not available in the training data set.
    }
    \label{fig7_20201205}
\end{figure*}

Next, let us assess how the current approach performs when the number of sensors are changed and when the sensors are in motion. 
We present the results for these cases in figure~\ref{fig7_20201205}. 
With $n_{\rm sensor}=100$ being the number of sensors observed during training, the reconstructed sea surface temperature field is in agreement with the reference field for both trained (left) and unseen sensor (middle) placements.
{Cases of unseen placements correspond to instances of sensors coming online or offline during development and being in motion.}
Despite that the input Voronoi tessellation being significantly modified with the displaced sensors, the reconstruction is still successful. 
This highlights the effective use of the mask image holding information on sensor locations. 
What is also noteworthy is that successful reconstructions can be achieved with $n_{\rm sensor}=70$ and $200$, which are numbers of sensors unseen during training. 
This corresponds to situations where sensors may come online or go offline during deployment.

The relationship between the number of input sensors and the $L_2$ error norm is also investigated.
The error level of the test data with unseen placements (orange curve) is higher than that with trained placements (blue curve) as shown in figure~\ref{fig7_20201205}.
{However, the present model achieves a reasonable estimation with an $L_2$ error being less than 0.1, even when the number of unseen sensors reaches 200.}
This result suggests that the present approach employing the Voronoi input and the mask image is robust for data sets where the number of sensors and the sensor placements vary considerably.
It also demonstrates the advantage of the present idea that a single trained model alone can handle the global field reconstruction for arbitrary number of sensors and time-varying positions.

\begin{figure*}[t]
    \centering
    \includegraphics[width=0.77\textwidth]{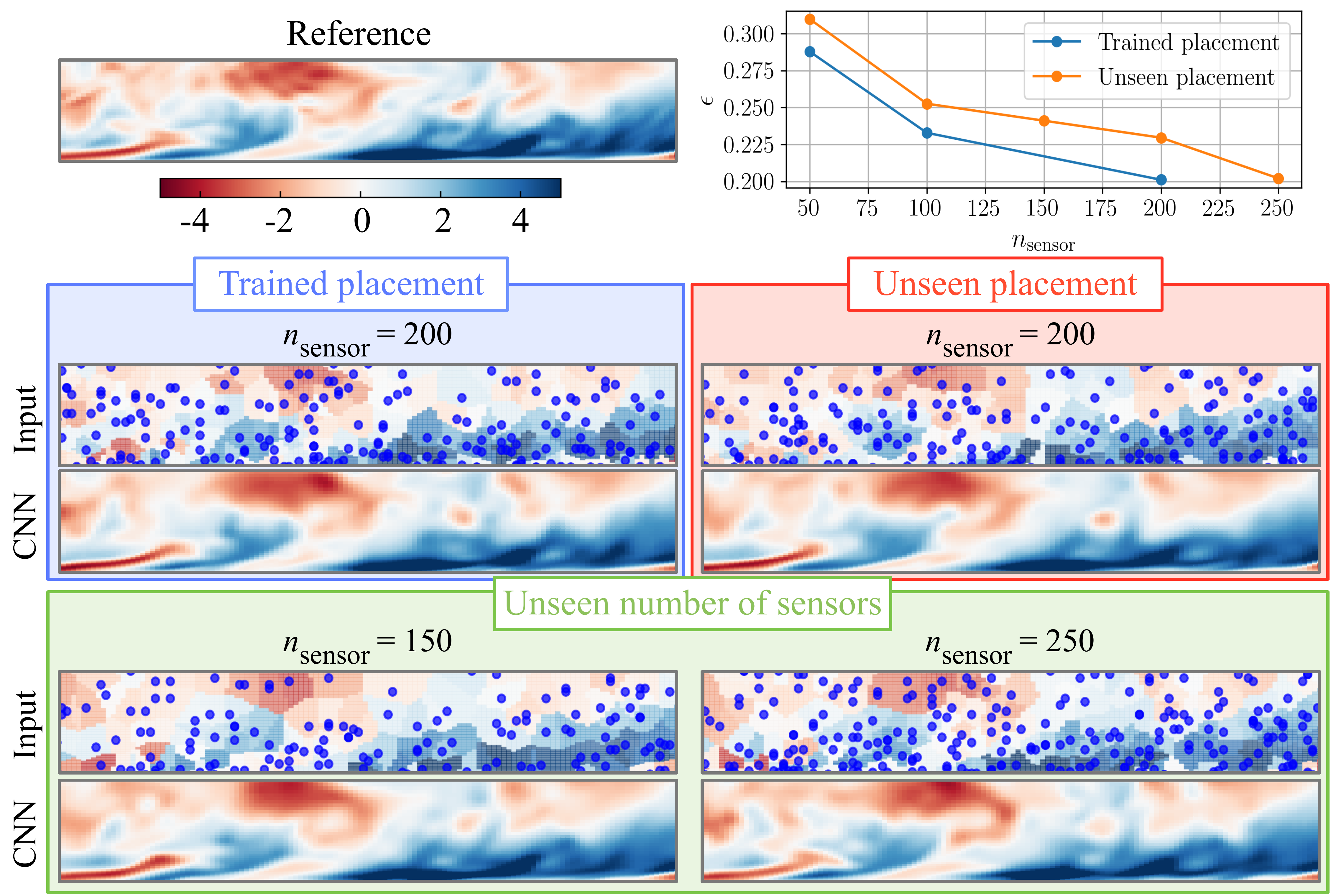}
    \caption{Voronoi tessellation-aided data recovery of turbulent channel flow.  Considered are $x-y$ cross sections of streamwise velocity fluctuation $u^\prime$ reconstructed with $n_{\rm sensor}=200$ (trained number of sensors) and $n_{\rm sensor}=\{150,250\}$ (untrained number of sensors).
    The error convergence over $n_{\rm sensor}$ is also shown. 
    }
    \label{fig7}
\end{figure*}

\subsection{Example 3: turbulent channel flow}
\label{sec:channel}

The above two problems contain strong periodicity in time, appearing as periodic vortex shedding and seasonal periodicity.
To further challenge the present approach, let us consider a chaotic and dynamically rich phenomenon of turbulent channel flow.
The flow field data is obtained by a direct numerical simulation of incompressible flow in a channel at a friction Reynolds number of $Re_{\tau}=180$.
Here, $x,y,$ and $z$ directions are taken to be the streamwise, wall-normal, and spanwise directions.
The size of the computational domain and the number of grid points are $(L_{x}, L_{y}, L_{z}) = (4\pi\delta, 2\delta, 2\pi\delta)$ and $(N_{x}, N_{y}, N_{z}) = (256, 96, 256)$, respectively, where $\delta$ is the half width of the channel.  
Details of the simulation can be found in Fukagata et al.~\cite{FKK2006}.
For the present study, $x-y$ section of a subspace is used for the training process, i.e., $x, y \in$ $[0, 2\pi\delta] \times [0, \delta]$ with $(N_x^*, N_y^*) = (128, 48)$.
The extracted subdomain maintains the same turbulent characteristics of the channel flow over the original domain, due to the symmetry of statistics in the $y$ direction and homogeneity in the $x$ direction~\cite{FFT2020b}.
We consider here the fluctuating component of an $x-y$ sectional streamwise velocity $u^\prime$ as the variable of interest.
For training, we use $n_{\rm snapshot}=10\,000$.
The numbers of sensors for training data are chosen to be $n_{\rm sensor, train}=\{50,100,200\}$ with 5 different cases of sensor placements.
For the test data, we also consider the use of 150 and 250 sensors with both trained and untrained sensor placements such that $n_{\rm sensor, test}=\{50, 100, 150, 200, 250\}$.
This setting allows us to assess the robustness of our approach for varied numbers of sensor inputs analogous to Example 2{, with consideration of both seen and unseen sensor locations for all numbers of sensors.}

The performance of Voronoi-assisted CNN-based spatial data recovery for turbulent channel flow is summarized in figure~\ref{fig7} for $n_{\rm sensor}=\{150, 200, 250\}$.
These numbers of sensors amount to 2.44\%, 3.26\%, and 4.07\% with respect to the number of grid points over the field.
We observe that finer flow features can be accurately reconstructed from just 200 sensors indicating a remarkable degree of sparsity in measurement.
Although error levels for unseen placement (i.e., the same number of sensors but at different locations) is higher than that for trained sensor placement, similar trends are obtained. 
Notably, reasonable reconstruction for both unseen number of sensors and unseen input sensor placement as shown in the results for $n_{\rm sensor}=\{150,250\}$. 
As these results suggest, the present approach is a powerful tool for global reconstruction of complex flow fields from sparse sensor measurements.

{

\begin{figure}[t]
    \centering
    \includegraphics[width=0.7\textwidth]{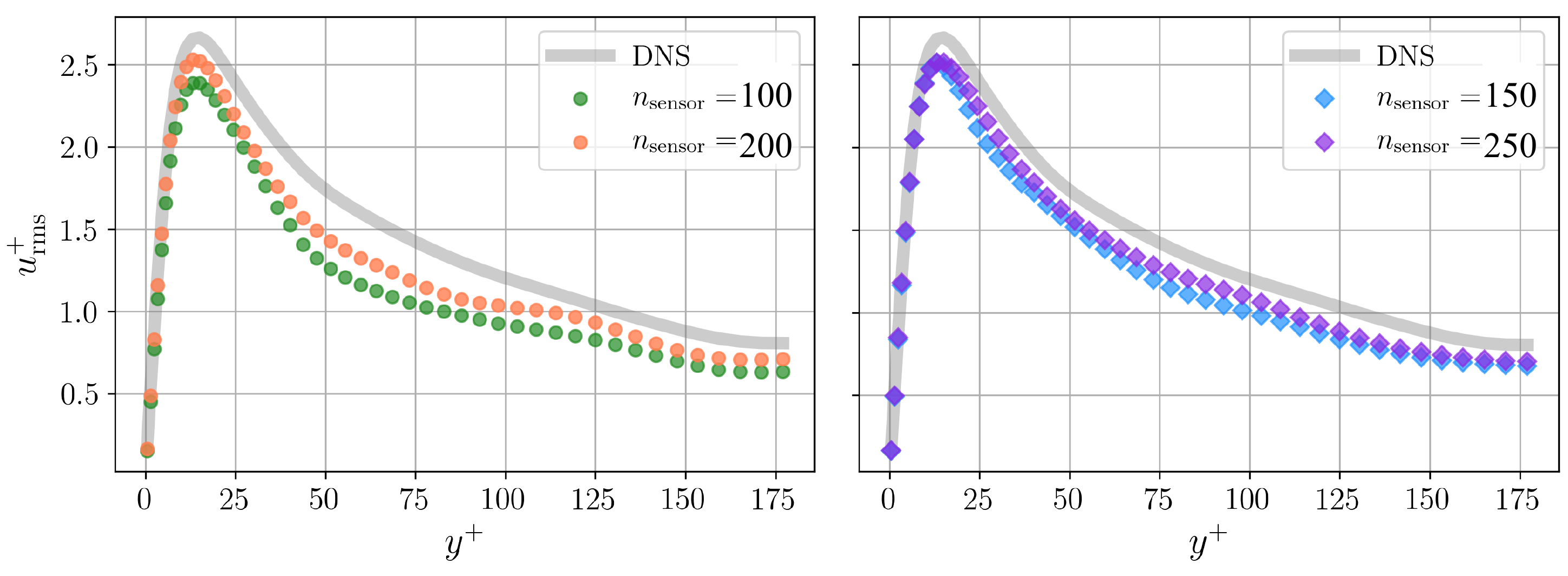}
    \caption{{Root mean squared value of streamwise velocity fluctuation in turbulent channel flow.
    }}
    \label{fig_R1}
\end{figure}

To quantitatively analyze reconstructed flow field, we examine the root-mean-square of the streamwise velocity fluctuations, as shown in figure~\ref{fig_R1}.
The peak in the near-wall region and profile are captured well for all considered numbers of sensors. 
The underestimations against the reference DNS curve are likely due to the use of fluctuation components and the use of an $L_2$ minimization during the training process~\cite{FNKF2019,FFT2020b}.
Since the supervised model minimizes a loss function through a training process, the machine learning model tends to provide solution near the average value of training data to reduce the loss function. 
This implies that the present machine learning model underestimates the fluctuations. 
Note that this is due to the supervised machine learning framework and is not an issue imposed by the unstructured grid placements or the use of Voronoi tessellation. 
This issue can be mitigated by using a loss function augmentation to the data, such as through $z$-scoring.
}

\begin{figure*}
    \centering
    \includegraphics[width=1\textwidth]{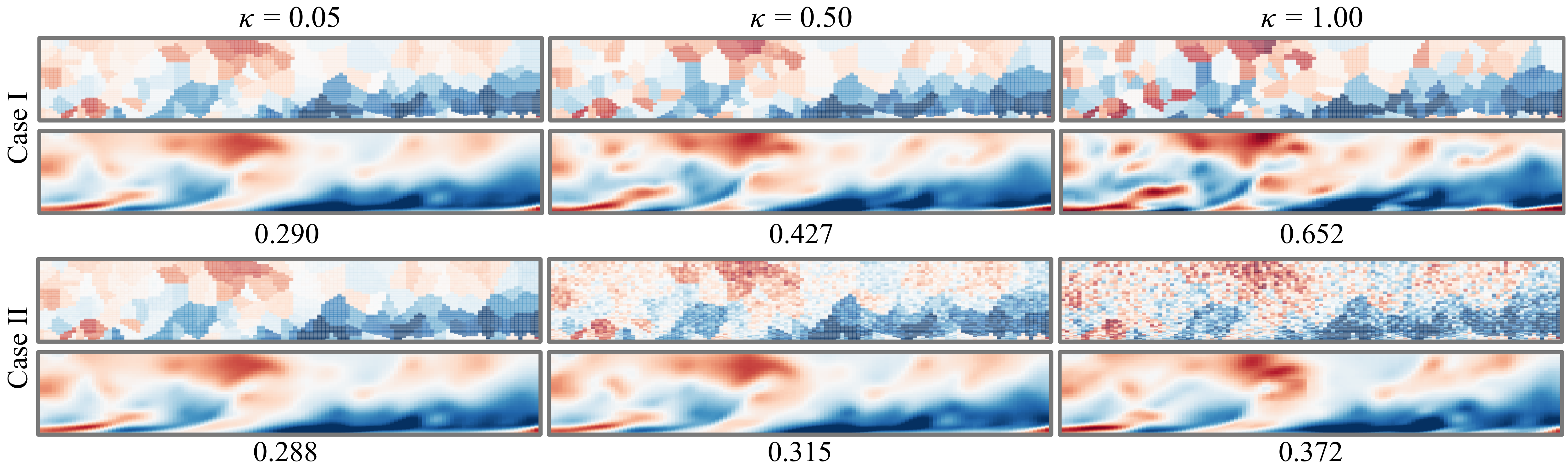}
    \caption{Robustness of the present reconstruction technique for noisy input for the example of turbulent channel flow. Input Voronoi tessellations (top) and reconstructed $x-y$ sectional flow fields (bottom) are shown for Cases I and II with $\kappa=0.05, 0.5,$ and $1.0$.
    Reference solution is shown in figure~\ref{fig7}.
    }
    \label{fig13}
\end{figure*}

To further assess the practical application of the present model, we analyze the effect of input noise on reconstruction. 
Here, we consider the influence of two types of perturbations to the training data.
Case I: Add noise to local sensor measurements ${\bm s}_m$ before performing the Voronoi tessellation; and Case II: Add noise to the Voronoi tessellation input ${\bm s}_V$.
For Cases I and II, the error is assessed as
\begin{align}
    &\epsilon_{\rm I}=||{\bm q}_{\rm ref}-{\cal F}({\bm s}_V({\bm s}+\kappa \bm n),{\bm s}_m)||_2/||{\bm q}_{\rm ref}||_2,\\
    &\epsilon_{\rm II}=||{\bm q}_{\rm ref}-{\cal F}({\bm s}_V+\kappa \bm n, {\bm s}_m)||_2/||{\bm q}_{\rm ref}||_2,
\end{align}
respectively, where $\kappa$ is the magnitude of the noise and ${\bm n}$ is the Gaussian distributed noise.

The reconstructed turbulent flows for $n_{\rm sensor}=200$ with the noisy inputs are summarized in figure~\ref{fig13}. 
As shown, the input Voronoi tessellations defined by Case II exhibit noisy features compared to those utilized by Case I since the noise for Case II is added after the preparation of the Voronoi tessellations.
The influence of noise for Case I is larger than that of Case II.
This is caused by the fact that the present CNN is trained for learning the relationship between the input Voronoi images and high-resolution flow fields, which implies that large perturbations to the sensor measurements produce greater error to the input images.
If robustness is desired, we recommend adding noise to the sensor measurements prior to the preparation of the Voronoi-based input images during the training process.
Overall, the present reconstruction technique is found to be robust against noise.

\section{Discussion}
\label{sec:conclusion}

Reconstruction of a global field variable from an arbitrary collection of sensors has been a longstanding challenge in engineering and the sciences.
In order to address this problem, we presented a data-driven global reconstruction technique comprised of Voronoi tessellation and CNN.
The present method relies on inputs of mask images holding the sensor location and Voronoi images representing the sensor measurements.  
The use of Voronoi tessellation translates the input sensor data to be represented on a uniform grid, which then enables the applications of CNNs to derive deep learning based reconstruction models.  
Three examples of global flow reconstruction from local sensor measurements demonstrated the accuracy and robustness of the current method.

Since CNNs have a large collection of toolsets from the image processing community, the present reconstruction approach is significant from the point of view of merging image processing with sensor data analytics. 
This perspective now allows engineers and scientists to move through the wealth of local and global measurements using data-driven techniques.   
Furthermore, Voronoi tessellation has the beneficial property of being able to discretize the spatial information in an adaptive manner only where sensor arrangements change in time.
This provides computational savings and an opportunity to develop spatially adaptive techniques.

{
The present approach can be extended to enforce physical constraints and properties~\cite{raissi2019physics,raissi2020hidden,HFMF2020a,LY2019}.
For example, evolutional deep neural network~\cite{du2021evolutional} has been able to enforce the divergence free constraint for incompressible flow.
Another possible extension is the construction of robust models with regard to physical parameters by preparing proper training data sets~\cite{morimoto2020generalization,HFMF2020b,kim2021unsupervised}.
To obtain generalizable models for turbulent data sets, unsupervised learning may also be helpful~\cite{kim2021unsupervised}.
}

The present data-driven approach was demonstrated with a set of local sensor measurements.
As this method performs spatial field reconstruction at each time, changes in the number of sensors or motion of the sensors are easily accommodated.  
Having this flexibility allows for extensions to incorporate other types of measurements, such as under-resolved satellite based measurements or particle image velocimetry.  
The current formulation also opens a path to incorporate intelligent sensor placements~\cite{manohar2018data} to further reduce reconstruction error and enhance the robustness with data redundancy.
{Moreover, the present method can be applied to cases where output attributes are different from the input data attributes.
In such a scenario, the Voronoi-tessellation is akin to a projection operation to extend learning one step further.
While we use the Voronoi CNN solely for flow reconstruction in the present paper, high-order turbulence statistics, e.g., root-mean-squared values of velocity fluctuations, can be extracted as well.}
The power and simplicity of the present approach will support scientific endeavor across a wide range of studies for complex data structures.

\section*{Data availability}

Training data used in the present study are available on Google Drive: 
Example 1 (two-dimensional cylinder wake): \url{https://drive.google.com/drive/folders/1K7upSyHAIVtsyNAqe6P8TY1nS5WpxJ2c?usp=sharing}),
Example 2 (NOAA sea surface temperature): \url{https://drive.google.com/drive/folders/1pVW4epkeHkT2WHZB7Dym5IURcfOP4cXu?usp=sharing}),
and Example 3 (turbulent channel flow):
\url{https://drive.google.com/drive/folders/1xIY_jIu-hNcRY-TTf4oYX1Xg4_fx8ZvD?usp=sharing}).

\section*{Code availability}

Sample codes for training the present models are available on GitHub (\url{https://github.com/kfukami/Voronoi-CNN}).

\bibliography{sample}

\section*{Acknowledgement}

R.M. and N.R. were supported by the U.S. Department of Energy, Office of Science, Office of Advanced Scientific Computing Research, under Contract~DE-AC02-06CH11357. 
This research was funded in part and used resources of the Argonne Leadership Computing Facility, which is a DOE Office of Science User Facility supported under Contract DE-AC02-06CH11357. 
Ko.F. thanks support from the Japan Society for the Promotion of Science (18H03758, 21H05007).
K.T. acknowledges support from the US Air Force Office of Scientific Research (FA9550-16-1-0650 and FA9550-21-1-0178) and the US Army Research Office (W911NF-19-1-0032).

\section*{Author contributions statement}

Ka.F., R.M, and N.R. designed research; Ka.F. performed research; Ka.F. analyzed data; Ka.F. and K.T. wrote the paper; Ko.F. and K.T. supervised.
All authors reviewed the manuscript.

\section*{Declaration of interest}

The authors report no conflict of interest.

\end{document}